\documentclass[12pt]{book}

\usepackage[dvips]{graphicx,color}
\usepackage{makeidx,view}
\usepackage{epsfig,psfig}

\makeauthorindex
\makeindex

\begin{document}

\BookTitle{\itshape The Universe Viewed in Gamma-Rays}
\CopyRight{\copyright 2002 by Universal Academy Press, Inc.}
\pagenumbering{arabic}

\chapter{
Chandra Studies of Nonthermal Emission from Supernova Remnants and
Pulsar Wind Nebulae}
\author{%
Patrick Slane\\
{\it Harvard-Smithsonian Center for Astrophysics}\\
{\it 60 Garden Street}\\
5A
{\it Cambridge, MA 02138, USA}}

%
\AuthorContents{R.\ Enomoto M.\ Mori, and S.\ Yanagita} 

\AuthorIndex{Enomoto}{R.}
\AuthorIndex{Mori}{M.} 
\AuthorIndex{Yanagita}{S.} 

%
%

\section*{Abstract}

While supernova remnants (SNRs) have long been considered prime candidates 
as sources of cosmic rays, it is only recently that X-ray observations have 
identified several shell-type SNRs dominated by nonthermal emission, thus 
revealing shock-accelerated electrons with energies extending far beyond 
the typical thermal spectrum. Two of these SNRs have been detected as sources
of VHE $\gamma$-rays.
In other remnants, discrepancies between the shock velocity 
and the electron temperature point to a strong cosmic ray component that 
has thrived at the expense of the thermal gas. Modeling of the radio, X-ray, 
and $\gamma$-ray emission provides constraints on particle acceleration 
as well as the properties of the medium in which the mechanism prospers.

Crab-like pulsar wind nebulae (PWNe) are characterized by a termination shock 
at which the wind is forced to join the slow expansion of the outer 
nebula. These shocks also act as sites in which particles are boosted to 
high energies; the X-ray emission from the Crab Nebula, as well as the inverse
Compton radiation observed as VHE $\gamma$-rays, imply electrons with energies
in excess of $\sim 100$ TeV. Recent X-ray observations have 
begun to reveal these 
shock zones in the Crab and other PWNe, and are now allowing us to constrain 
the nature of pulsar winds as well as the flow conditions in the outer 
nebulae. Here I present a brief overview of recent studies with the 
{\em Chandra X-ray Observatory} in which the properties of these shock 
acceleration regions are finally being revealed.

\section{Introduction}

Supernova remnants and their associated pulsars are sites in which strong 
shocks act to accelerate particles to extremely high energies. 
The connection between SNRs and 
the energetic cosmic rays that pervade the Galaxy has long been assumed, 
for example; shock acceleration by the SNR blast wave provides ample 
energy for the production of multi-TeV particles, and recent observations
of nonthermal X-ray and VHE $\gamma$-ray emission from several
SNRs has confirmed the presence of electrons at these high energies.
At the same time, models for the structure of PWNe predict particle 
acceleration at the wind termination shock, and recent X-ray observations
have begun to probe these acceleration sites.  These X-ray measurements
provide constraints on the expected higher energy emission processes, and
are thus of considerable interest in the context of VHE $\gamma$-ray
astronomy.  Here I review recent {\em Chandra} studies of nonthermal 
X-ray emission in SNRs and PWNe, with particular emphasis on the evidence
for particle acceleration to very high energies.

\section{Nonthermal X-rays and Particle Acceleration in SNRs}

The radio synchrotron emission from SNRs is testimony to the presence
of energetic particles. However, this corresponds to electron energies 
far below the ``knee'' of the cosmic ray spectrum at $\sim 10^{15}$~eV:
\begin{equation}
E_{\rm GeV} \approx \left[\frac{\nu}{16 {\rm\ MHz}} B_\mu^{-1}\right]^{1/2}
\end{equation}
where $\nu$ is the frequency of the radio emission
and $B_\mu$ is the magnetic field strength in $\mu$G. 
X-ray observations allow us to probe much higher energy particles, and
also characterize the thermodynamic states of SNRs and infer the 
dynamics of their evolution.
For an ideal gas, the thermal postshock temperature is
\begin{equation}
T = \frac{3 \mu m}{16 k} V_s^2
\end{equation}
where $V_s$ is the shock speed, $m$ is the proton mass, and $\mu$ is
the mean molecular weight of the gas ($\mu \approx 0.6$).
This shock-heated gas yields the familiar X-ray emission, characterized
by a thermal bremsstrahlung spectrum accompanied by strong emission lines.

In addition to thermal heating of the swept-up gas, some fraction of the
shock energy goes into nonthermal production of
relativistic particles through diffusive shock acceleration.
The maximum particle energy in such a scenario can be limited by radiative
losses, the finite age of the SNR, or particle escape from the accelerating
region. If the relativistic particle component of the
energy density becomes comparable to that of the thermal component,
the shock acceleration process can become highly nonlinear. The gas
becomes more compressible, which results in a higher density and
enhanced acceleration. 
When the acceleration is very efficient, the relationship between the
shock velocity and the mean postshock temperature is no longer well
approximated by Eq. 2; the acceleration process depletes 
thermal energy, and the temperature for a given shock velocity drops below
that expected in the test particle case (Decourchelle et al. 2002).
This process has been considered in detail
by Baring et al. (1999) who present a model for the radio to
$\gamma$-ray emission. 
The broadband spectra of SNRs depend highly on ambient conditions,
and X-ray studies of these SNRs reveal these conditions and can provide
spectral measurements which strongly constrain the models. 

In the simplest picture, the passage of material through the SNR shock results
in electrons and ions being boosted to the velocity of the shock. Because of
the mass difference, this means that the electrons and ions are not initially
in temperature equilibrium. The maximum timescale for equilibration is that
provided by Coulomb interactions, but plasma processes may reduce this
considerably. The state of equilibration is important, because while
the dynamics of SNR evolution are dominated by the ions (which carry the bulk
of the momentum), it is the electrons that produce the X-ray emission we
observe. Thus, when temperature measurements are used to infer the shock
velocity, for example, the state of temperature equilibration is exceedingly
important. 

As an example, the blast wave
speed inferred from an expansion
study comparing a {\em Chandra} image of 1E~0102.2--7219 (Figure 1, left),
with high resolution images
taken with {\it Einstein} and {\it ROSAT} indicates a post-shock temperature
which is much higher than the observed electron temperature (Hughes et al. 
2001). In this case, the discrepancy appears
to be larger than can be accounted for assuming Coulomb equilibration,
suggesting that a significant fraction of the shock energy has gone into
cosmic ray acceleration rather than thermal heating of the postshock gas.
Higher resolution X-ray expansion studies of 1E~0102.2--7219 (and other
young SNRs) are needed to
confirm this scenario, but the notion that particle acceleration is
efficient enough in some SNRs to significantly affect their dynamics
has strong foundations from recent studies of other SNRs. As we discuss
below, direct evidence of very energetic electrons now exists for a
handful of shell-type SNRs. In addition to the two SNRs which we
discuss in the following sections (and also G266.2--1.2; Slane et al. 
2001), for which synchrotron radiation dominates 
the X-ray emission, evidence for energetic particles in the form of 
nonthermal filaments or hard
tails in the X-ray spectra have also been observed for Cas~A~
(Allen et al. 1997, Hughes et al. 1999, Vink et al. 1999),
RCW~86 (Borkowski et al. 2001, Rho et al. 2002),
and other SNRs. 

\begin{figure}[tb]
\centerline{\psfig{file=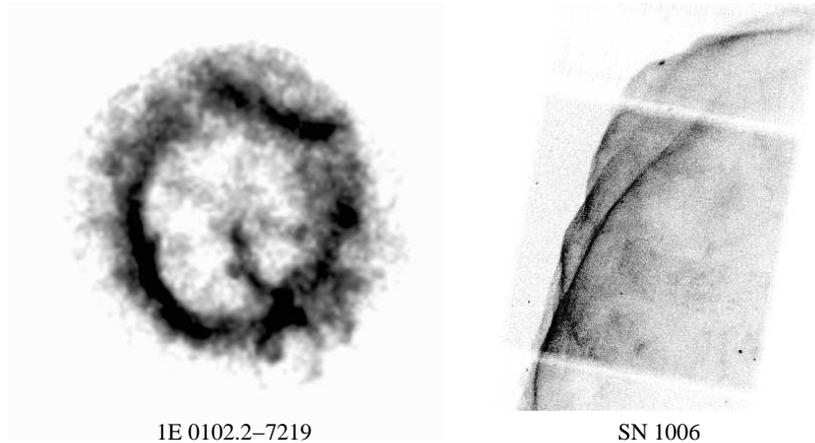,height=2.3in}}
\caption{
Left: {\em Chandra} image of 1E~0102.2--7219, a young oxygen-rich SNR in
the SMC. A comparison of the expansion rate and shock temperature indicates
that considerable energy has gone into particle acceleration.
Right: {\em Chandra} image of the northeast limb of SN~1006. The diffuse
X-ray emission in the interior regions is thermal emission from the hot
postshock gas. The outer limb, which is observed to consist of bright
filaments, is dominated by synchrotron from extremely energetic electrons.
}
\end{figure}

\subsection{SN~1006}
The first evidence of multi-TeV particles directly associated with 
a shell-type SNR was uncovered in studies of SN~1006 with the ASCA 
observatory (Koyama et al. 1995). While spectra from the central 
regions of the SNR show distinct line emission associated with shock-heated
gas, the emission from the bright limbs of the remnant was shown to be
completely dominated by synchrotron emission. Reynolds (1998) modeled 
the emission as the result of diffusive shock acceleration in a low
density medium with the magnetic field orientation providing the distinct
``bilateral'' morphology. Subsequent observations with the CANGAROO 
telescope revealed VHE $\gamma$-ray emission from one limb of the 
SNR (Tanimori et al. 1998),
thus confirming the presence of extremely energetic particles. The low
ambient density of this remnant, which resides well above the Galactic
Plane, is insufficient to explain the $\gamma$-ray emission as the result
of $\pi^0$ decay from proton-proton collisions. Rather, the emission
results from inverse-Compton scattering of microwave background photons
off the energetic electron population in SN~1006 (but see Berezhko --
these proceedings -- for an alternative picture). Using joint spectral
fits to the radio as well as thermal and nonthermal X-ray emission, 
Dyer et al. (2001) conclude that the total energy in relativistic
particles is $\sim 100$ times the energy in the magnetic field, confirming
the notion that the nonthermal particle component contributes 
significantly to the dynamics of the blast wave evolution.

{\em Chandra} observations of SN~1006 (Figure 1, right) reveal sharp
filamentary structure along the SNR rim. Diffuse emission beyond these
filaments is observed, presumably indicating the upstream diffusion
of particles accelerated in the shocks. The upstream scale length of these
nonthermal filaments appears to be considerably smaller than 
expected for standard
diffusive shock acceleration, however (Bamba et al. -- these proceedings), 
possibly indicating an upstream field nearly parallel to the plane of the
shock and a downstream magnetic field strength ($\sim 50 \ \mu{\rm
G}$) considerably higher than that previously estimated (e.g.
Reynolds 1998). A program to map SN~1006 in its entirety with 
{\em Chandra} is underway. This will form a baseline for future 
expansion measurements that will bear heavily on these questions.

\subsection{G347.3--0.5 (RX J1713.7-3946)}

\begin{figure}[tb]
\centerline{\psfig{file=fig2a.ps,height=2.3in}
\psfig{file=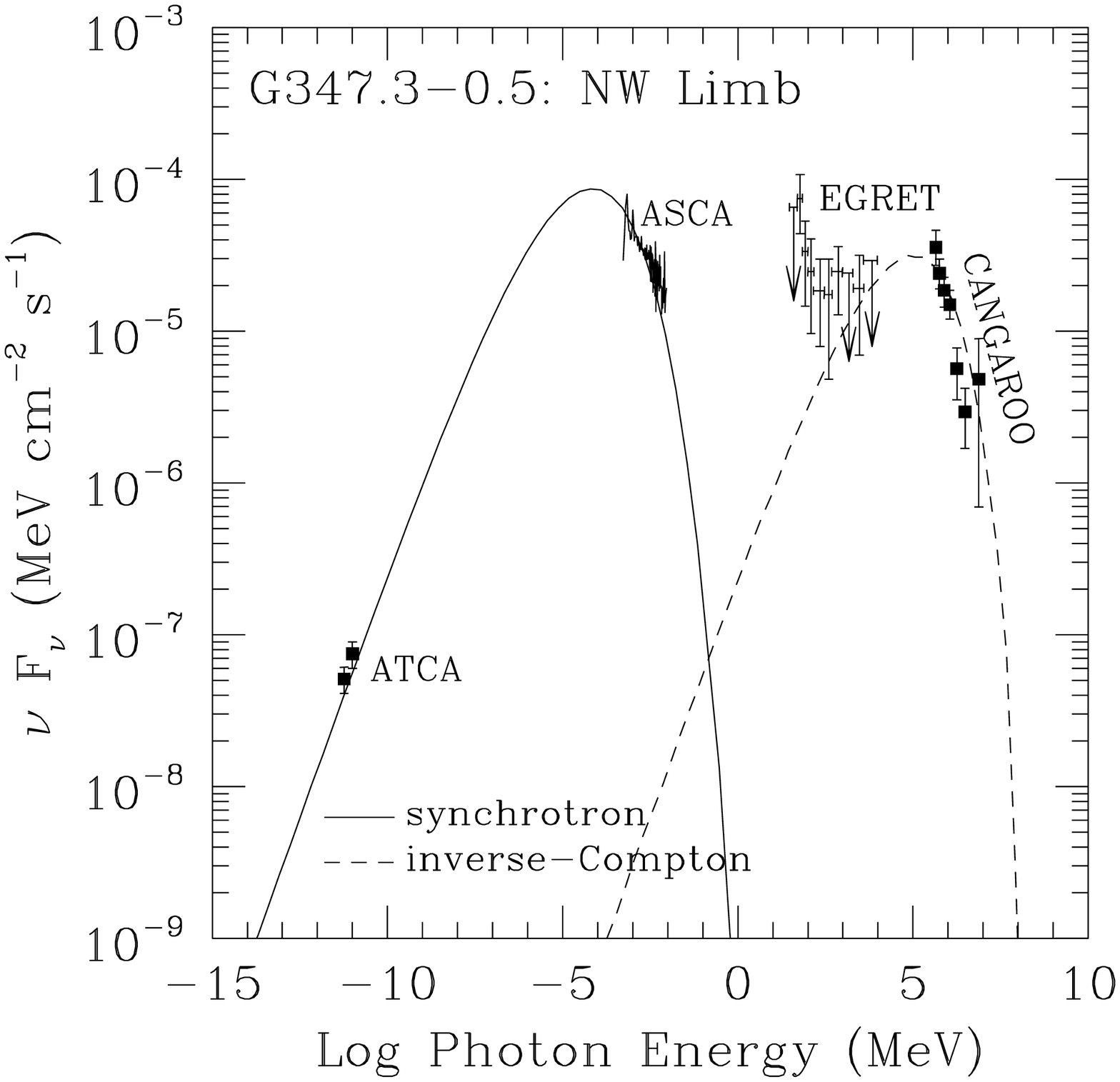,height=2.3in}}
\caption{
Left: Radio contours of G347.3--0.5 from the ATCA, along with 
{\em Chandra} images of the NW and SE regions of the SNR. X-ray
filaments in the NW follow radio arcs; VHE $\gamma$-ray emission is
detected from this region. Right: Model of the broad-band emission from 
G347.3--0.5 assuming a power law distribution of electrons with an
exponential cutoff ($E_{\rm max} \sim 3$~TeV) and a magnetic filling 
factor comprising only $\sim 0.7\%$ of the inverse-Compton emitting 
region (Lazendic et al. 2003).
}
\end{figure}

ASCA observations of G347.3--0.5 (Koyama et al. 1997, Slane et al. 1999)
established this as the second member of the class of shell-type SNRs for
which the X-ray flux is dominated by synchrotron radiation. Unlike 
SN~1006, this SNR appears to have evolved in the vicinity of dense
molecular clouds (Slane et al. 1999) with which the shock may now be
interacting. CANGAROO observations (Muraishi et al. 2000)
reveal VHE $\gamma$-ray emission from 
the vicinity of the northwest rim, which is brightest in X-rays. Combining
radio measurements from the ATCA with the X-ray and $\gamma$-ray 
results, Ellison, Slane, \& Gaensler (2001)
used diffusive shock acceleration models to
conclude that the radio and X-ray emission results from synchrotron radiation 
from a nonthermal electron population accelerated by the forward shock, and 
that the $\gamma$-ray
emission can be self-consistently modeled as inverse-Compton emission.
Combined with limits on the ambient density based on the lack of thermal
X-ray emission, the models indicate very efficient particle acceleration
with $> 25\%$ of the shock kinetic energy going into relativistic 
ions. A comparison of {\em Chandra} observations of the northwest rim with
high resolution radio maps from the ATCA 
shows good overall agreement with the radio and X-ray morphology in this 
region (Figure 2, left), consistent with the interpretation that the emission
comes from the same electron population (Lazendic, et al. 2003).

The nearby unidentified EGRET source 3EG J1714-3857 has been suggested as
being associated with G347.3--0.5, possibly 
resulting from the decay of neutral pions produced in the collision
of accelerated ions with the nearby molecular clouds (Butt, et al. 2001), 
although this
would appear inconsistent with the X-ray measurements unless the emission
originates from a distinct spatial region. Most recently, Enomoto et 
al. (2002) have presented new CANGAROO data which, they argue, 
establish $\pi^0$-decay
as the mechanism by which the TeV $\gamma$-rays are produced. However,
the predicted spectrum over-predicts emission from the EGRET band by a
large margin (Reimer et al. 2002, Butt et al. 2002), making the claim for 
direct observation of ion acceleration appear problematic.

The CANGAROO spectrum can be modeled as inverse-Compton emission from the same 
electron spectrum that produces the synchrotron radiation, while still falling 
below the EGRET limits, if the strong magnetic field region in which the 
synchrotron emission comprises only a small fraction of the total volume
over which the inverse-Compton emission is produced (Figure 2, 
right; Lazendic et al. 2003). The resulting magnetic field filling factor is
only $\sim 0.7\%$, which is quite small although the long diffusion lengths
for the energetic particles certainly make the inverse-Compton emitting 
region considerably larger than that of the compressed magnetic field region
where the bulk of the synchrotron emission arises. The field strength in 
the synchrotron filaments
must be $\sim 50 \mu{\rm G}$, which is very high, though similar to values
found in maser-producing regions of molecular clouds. 
Uchiyama et al. (2002) have also modeled the synchrotron emission in G347.3--0.5
to assess the required magnetic field conditions. For a distance
of 6 kpc, they derive $B \sim 50 \mu{\rm G}$ -- similar to 
that obtained in our analysis -- but require that this field be
similar in the filamentary and diffuse emission regions. This leads to
the conclusion that the spectral cutoff in the diffuse emission regions
is due to radiative losses while that in the filaments is due to 
diffusive escape of the particles. The resulting inverse-Compton
emission falls far below the observed CANGAROO flux, thus
requiring some other mechanism to produce the TeV $\gamma$-ray emission.

\section{Energetic Particles from Pulsars and Their Wind Nebulae}

Our basic understanding of ``Crab-like'' SNRs stems from the picture presented
by Rees \& Gunn (1974), and expanded upon by Kennel \& 
Coroniti (1984a,b)
in which an energetic wind is injected from a
pulsar into its surroundings. 
As illustrated schematically in Figure~3,
the basic structure of a pulsar wind nebula is regulated by the input
power from the pulsar and the density of the medium into which the nebula
expands; the pulsar wind inflates a magnetic bubble which
is confined in the outer regions by the expanding shell of ejecta or
interstellar material swept up by the SNR blast wave. The boundary condition
established by the expansion at the nebula radius $r_N$ results in the
formation of a wind termination shock at which the highly relativistic
pulsar wind is decelerated to $v \approx c/3$ in the postshock region,
ultimately merging with the particle flow in the nebula. The
shock forms at the radius $r_w$ at which the ram pressure of the wind is
balanced by the internal pressure of the pulsar wind nebula:
\begin{equation}
r_w^2 = \dot E/(4 \pi \eta c p)
\end{equation}
where $\dot E$ is the rate at which the
pulsar injects energy into the wind, $\eta$ is the fraction of a spherical
surface covered by the wind, and $p$ is the total pressure outside the shock.

\begin{figure}[h]
\centerline{\psfig{file=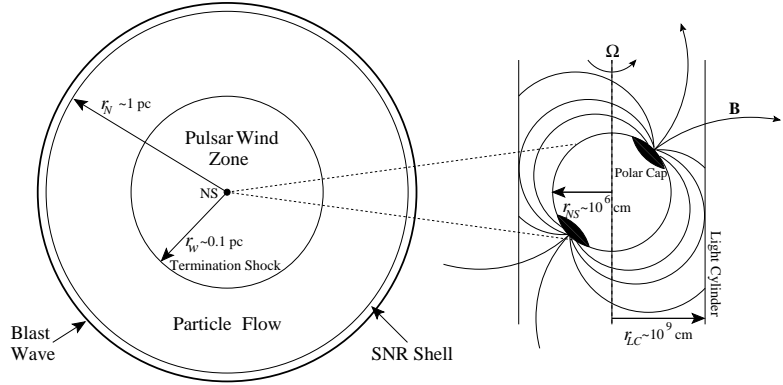,height=2.in}}
\caption{Schematic view of a pulsar and its wind nebula. See the text for
a complete description. (Note the logarithmic
size scaling in the PWN figure when comparing with images shown elsewhere in
the text.)}
\end{figure}

The dynamics of the particle flow yield $\gamma \sim 10^6$ for the electrons
in the postshock region (Arons \& Tavani 1994).
However, for typical magnetic
field strengths the observed X-ray emission requires $\gamma > 10^8$.
Particle acceleration at the termination shock apparently boosts the
energies of the wind particles by a factor of 100 or more, to energies
in excess of $\sim 50$~TeV. Arons \& Tavani (1994; see also 
Arons 2002)
note that this process cannot proceed by normal diffusive shock acceleration
because the magnetic field at the termination shock must be nearly
perpendicular to the flow. Rather, they argue that the e$^\pm$ acceleration
is the result of resonant cyclotron absorption of low frequency electromagnetic
waves emitted by ions gyrating in the compressed $B$-field of the hot
post-shock gas. 
Ultimately, the pressure in the nebula is believed to reach the equipartition
value; a reasonable pressure estimate can be obtained by integrating
the radio spectrum of the nebula, using standard synchrotron emission
expressions, and assuming equipartition between particles and the magnetic
field. Typical values yield termination shock radii of order 0.1~pc, which
yields an angular size of several arcsec at distances of a few kpc. 

The pulsar wind can be characterized in terms of $\dot E$
and the parameter parameter $\sigma$ representing
the ratio of Poynting flux to particle flux. For the Crab Nebula,
Kennel \& Coroniti (1984a) find
that small values of $\sigma$ ($\sim 10^{-3}$) are required, indicating a
particle-dominated wind. Yet current understanding of pulsar outflows
predicts $\sigma \gg 1$ where the wind is launched (Arons 2002). Somehow,
between the pulsar light cylinder and the wind termination shock the 
balance between the electromagnetic energy and the kinetic energy of the
flow changes dramatically. The ability to identify the termination shock
and measure the emission parameters in the immediate region is now 
providing constraints on the nature of the wind that may ultimately
help unravel this problem. Below I summarize recent X-ray investigations
that have finally begun to probe this important shock region in which
particle acceleration is apparently taking place.

\begin{figure}[t]
\centerline{\psfig{file=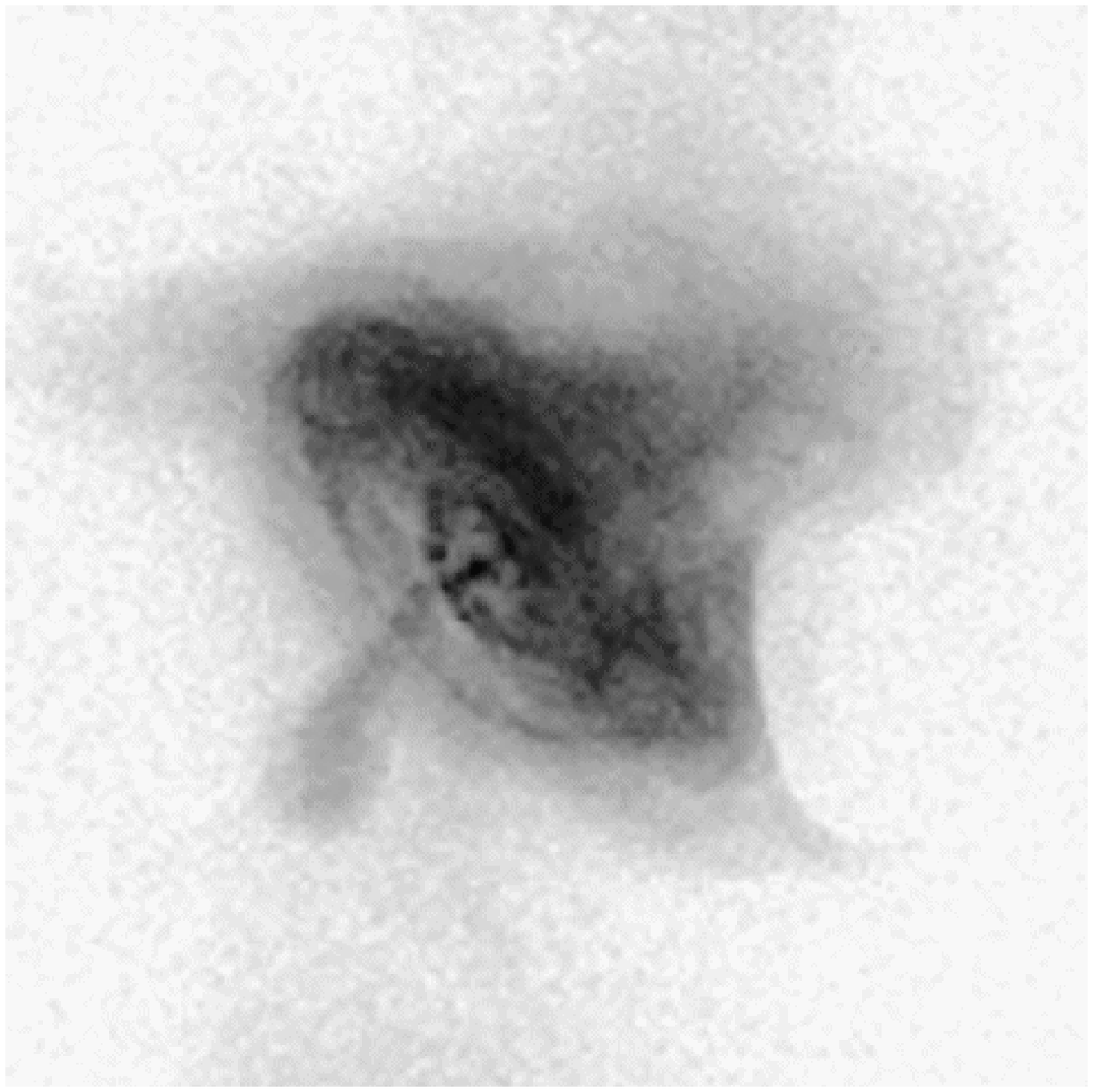,height=2.3in}
\psfig{file=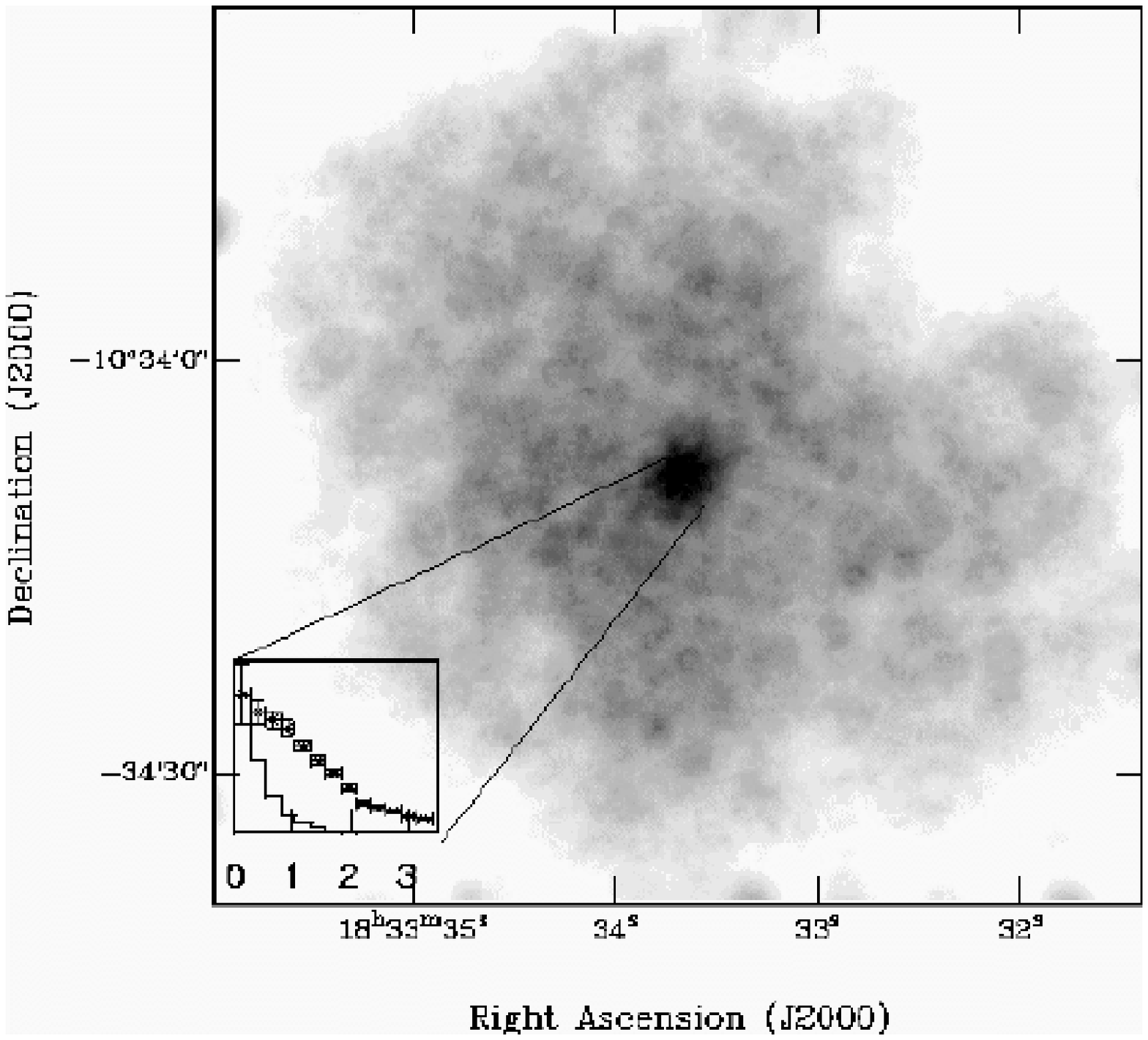,height=2.3in}}
\caption{
Left: {\em Chandra} image of the Crab Nebula showing the toroidal structure,
the jet, and the inner ring corresponding to the pulsar termination shock.
Right: {\em Chandra} image of G21.5-0.9, a PWN whose pulsar has not yet been
detected. The bright central X-ray source is resolved in this image (see
inset); the $\sim 2-3$ arcsec extent is consistent with the expected size
of the termination shock zone.
}
\end{figure}

\subsection{Crab Nebula}

The Crab Nebula is the best known of the class of pulsar wind nebulae 
and has inspired much of the theoretical work on PWNe. Powered
by an energetic central pulsar, it emits synchrotron radiation from radio
wavelengths up beyond the hard X-ray band. Optical wisps are observed
in the inner nebula, at a position interpreted as the pulsar wind termination
shock (Hester 1995), and high resolution X-ray observations 
(Figure 4, left) reveal a distinct ring of emission in this same region 
as well as a jet emanating from the pulsar (Weisskopf et al. 2000).
Moreover, monitoring observations of the Nebula (Hester 1995, Mori et al.
2002)
show that these and other detailed features are highly dynamic. 
The discovery of radio wisps in inner ring region (Bietenholz et al. 2002)
suggests that the acceleration site may be the same for the entire population
of electrons that produce the broad-band synchrotron emission. 

In addition to the jet and inner ring, the X-ray image reveals an outer
toroidal structure that presumably lies in the equatorial plane, as well as
fine structure correlated with optical polarization measurements, indicating 
that the structures trace the magnetic field. The early models
of Rees \& Gunn (1974) and Kennel \& Coroniti (1984a,b)
predict these basic
properties as the result of a wound-up magnetic field, the large-scale
confinement of the Nebula by the (unseen) supernova ejecta, and the termination
of the pulsar wind flow by an inner shock. This picture leads to the
inference of a
low-$\sigma$ wind described above. As we describe below, recent observations
of PWNe have begun to show many similar features, indicating that our basic
picture -- while still poorly understood in detail -- can at least be said
to apply to a ``class'' of objects.

\subsection{G21.5--0.9}

G21.5--0.9 is a compact SNR that exhibits strong linear
polarization, a flat spectrum, and centrally peaked emission in the radio band.
The SNR has a lower $L_x/L_r$ ratio than the Crab; it is a
factor of $\sim 9$ less luminous in the radio and a factor of $\sim 100$
less in X-rays.
To date, there has been no detection of a central pulsar. However,
{\em Chandra}\ observations by Slane et al. (2000)
reveal a compact source
of emission at the center of the remnant as well as a
radial steepening of the spectral index consistent with synchrotron burn-off
of high energy electrons injected from a central source.
Using an empirical relationship between the total X-ray luminosity
of the PWN with the spin-down power of the pulsar powering the 
nebula (Seward \& Wang 1988) suggests
the presence of a pulsar with $\dot E = 10^{37.5}{\rm\ erg\ s}^{-1}$, although
Chevalier (2000)
argues that the spectral variations in the nebula imply
more efficient conversion of $\dot E$ into X-ray emission, and suggests that
$\dot E \approx 10^{36.7}{\rm\ erg\ s}^{-1}$ is more likely for the pulsar in
this PWN. Detection of pulsations from the central source are required to
address this further.
Using the larger $\dot E$ estimate along with pressure estimates from
the radio spectrum, Eq. 3 predicts a wind termination shock at a
radius of $\sim 1.5 \eta^{-1/2}$~arcsec from the pulsar, 
assuming a distance of 5~kpc. As indicated in the inset to Figure~4, 
the brightness profile of the compact X-ray source in G21.5--0.9
is broader than that for a point source. The $\sim 2^{\prime\prime}$ extent
of the source is consistent with the expected size of the 
termination shock zone. 

Slane et al. (2000) also report the presence of 
a faint extended shell of
emission surrounding G21.5--0.9 whose featureless spectrum may be associated
with energetic particles accelerated by the SNR blast wave, similar to those
observed for SN~1006 and other SNRs as described in Section 2. 

\subsection{PSR J1124-5916/G292.0+1.8}
\begin{figure}[t]
\centerline{\psfig{file=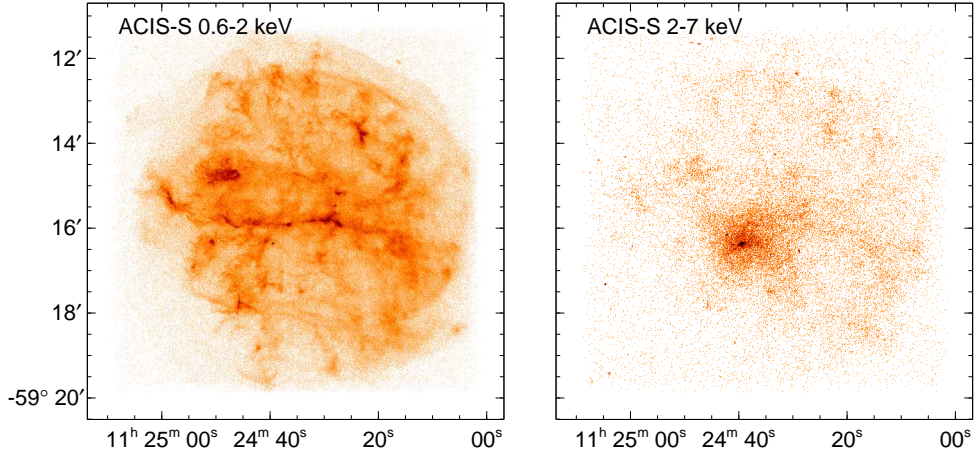,angle=90,height=2.3in}}
\vspace{0.1in}
\caption{
{\em Chandra} ACIS-S X-ray images of G292.0+1.8
in soft (left: 0.6--2 keV) and hard (right: 2--7 keV) X-ray bands. The
hard band shows a diffuse, plerionic nebula containing a
central point-like source.
}
\end{figure}


G292.0+1.8 is a member of the ``oxygen-rich'' class of SNRs with optical
emission revealing the presence of ejecta whose tell-tale composition points
to a massive star progenitor. Previous studies have shown that the X-ray 
emission is dominated by ejecta, and have revealed a region of hard emission
indicative of a pulsar-driven nebula. {\em Chandra} observations 
reveal an unprecedented example of a composite SNR
showing both such components (Hughes et al. 2001, Park et al. 2002).
Figure 5 (left) shows a soft X-ray image of the remnant.
The bulk of the interior emission is dominated by ejecta, and spectra from
discrete regions suggest large-scale variations in
ionization state and elemental abundance within the ejecta.  
The bar-like
structure running across the center, along with the thin filamentary
structure encircling most of the SNR, has solar abundances indicative
of material swept up by the forward shock.
The hard-band image in Figure 5 (right) reveals a 
neutron star with a surrounding pulsar wind nebula, similar in size
to the Crab. Recent radio observations (Camilo et al. 2002) have identified 
a young pulsar (PSR J1124-5916) with a 135~ms rotation period that appears 
to be the counterpart. The {\em Chandra} image of the compact X-ray source
is marginally extended, with a size consistent with that expected for the
termination shock zone of the pulsar wind.

\section{Summary}

The strong shocks in SNRs and PWNe have long been regarded
as sites where particles are accelerated to extremely high energies. SNRs,
in particular, are leading candidates as the source of cosmic rays up to
the knee of the spectrum. X-ray emission from PWNe also require particles
with higher energies than expected in the postshock flow of the pulsar 
wind, indicating efficient acceleration at the termination shock. X-ray
observations have now begun to provide direct evidence of the energetic
particles and shock structures where this such acceleration takes place.
Through studies of the dynamics and nonthermal X-ray emission from SNRs,
and of the wind termination shocks and associated particle outflows in 
PWNe, strong constraints are being placed on models of the particle 
acceleration process. High resolution
X-ray observations promise continued advances in this area, through
measurements of SNR nonthermal emission and expansion rates, and the
inner structure of PWNe. In addition, $\gamma$-ray observations with
current and upcoming \v{C}erenkov telescopes, as well as future
space-borne observatories, hold considerable promise for probing these
sites of extremely energetic particles.

\section*{Acknowledgments}
I would like to thank Bryan Gaensler, Jack Hughes, Don Ellison,
Jasmina Lazendic, and Steve Reynolds for their contributions as collaborators 
on much of the above work. I also thank the organizers of this meeting
for their kind invitation to present this work. This 
research was funded in part by NASA Contract NAS8-39073 and Grants 
NAG5-9281 and GO0-1117A.

\section*{References}
\re
1.\ Allen, G. E. et al. 1997, ApJ, 487, L97
\re
2.\ Arons, J. 2002, in ``Neutron Stars in Supernova Remnants,''
ASP Conference Series, Vol. 271, eds P. O. Slane and B. M. Gaensler, p. 71
\re
3.\ Arons, J. \& Tavani, M. 1994, ApJS, 90, 797
\re
4.\ Baring, M. G. et al. 1999, ApJ, 513, 311
\re
5.\ Bietenholz, M. F., Frail, D. A., \& Hester, J. J. 2001, ApJ, 560, 254
\re
6.\ Borkowski et al. 2001, ApJ, 550, 334
\re
7.\ Butt, Y., et al. 2001, ApJ, 562, L167
\re
8.\ Butt, Y., et al. 2002, Nature, 418, 499
\re
9.\ Camilo et al. 2002, ApJ, 567, L71
\re
10.\ Chevalier, R. A. 2000, ApJ, 539, L45
\re
11.\ Decourchelle, A., Ellison, D. C., \& Ballet, J. 2002, ApJ, 543, L57
\re
12.\ Dyer, K. K. et al. 2001, ApJ, 551, 439
\re
13.\ Ellison, D. C., Slane, P., \& Gaensler, B. M. 2001, ApJ, 563, 191
\re
14.\ Enomoto, R. et al. 2002, Nature, 416, 823
\re
15.\ Hester, J. J. et al. 1995, ApJ, 448, 240
\re
16.\ Hughes, J. P. et al. 1999, ApJ, 528, L109
\re
17.\ Hughes, J. P. et al. 2001, ApJ, 559, L153
\re
18.\ Hughes, J. P., Rakowski, C. E.  \& Decourchelle, A. 2001, ApJ, 543, L61
\re
19.\ Kennel, C. F. \& Coroniti, F. V. 1984a, ApJ, 283, 694 
\re
20.\ Kennel, C. F. \& Coroniti, F. V. 1984b, ApJ, 283, 710
\re
21.\ Koyama, K. et al.  1995, Nature, 378, 255
\re
22.\ Koyama, K. et al. 1997, PASJ, 49, L7
\re
23.\ Lazendic, J. et al. 2003, ApJ, submitted
\re
24.\ Mori, K. et al. 2002, in ``Neutron Stars in Supernova Remnants,''
ASP Conference Series, Vol. 271, eds P. O. Slane and B. M. Gaensler, p. 157
\re
25.\ Muraishi, T. et al. 2000, A\&A, 354, L57
\re
26.\ Park et al. 2002, ApJ, 564, L39
\re
27.\ Rees, M. J. \& Gunn, R. E. 1974, MNRAS, 167, 1
\re
28.\ Reimer, O. et al. 2002, A\&A, 390, L43
\re
29.\ Reynolds, S. P. 1998,  ApJ, 493, 375
\re
30.\ Seward, F. D. \& Wang, Z.-R. 1988, ApJ, 332, 199
\re
31.\ Slane, P. et al. 1999, ApJ, 525, 357
\re
32.\ Slane, P. et al. 2000, ApJ, 533, L29
\re
33.\ Slane, P. et al. 2001, ApJ, 548, 814
\re
34.\ Tanimori, T. et al. 1998, ApJ, 497, L25
\re
35.\ Vink, J. et al. 1999 A\&A, 344, 289
\re
36.\ Weisskopf, M. C. et al. 2000, 536, L81

\endofpaper
\end{document}